\documentclass{article}
\usepackage{graphicx,amssymb}

\begin{document}
\title{Evidence for criticality in financial data}
\author{G. Ruiz L\'{o}pez
and A. Fern\'{a}ndez de Marcos\thanks{E-mail address: guiomar.ruiz@upm.es}\\
\it{Universidad Polit\'{e}cnica de Madrid }\\
\it{Pza. Cardenal Cisneros 3, 28040 Madrid, Spain}
}
\maketitle

\begin{abstract}
We provide evidence that cumulative distributions of absolute normalized returns for the $100$ American companies with the highest market capitalization, uncover a  critical behavior for different time scales $\Delta t$. Such cumulative distributions, in accordance with a variety of complex --and financial-- systems, can be modeled by  the cumulative distribution functions of  $q$-Gaussians, the distribution function that, in the context of nonextensive statistical mechanics, maximizes a non-Boltzmannian entropy. These $q$-Gaussians are characterized by two parameters, namely $(q,\beta)$, that are uniquely defined by $\Delta t$. From these dependencies, we find a monotonic relationship between $q$ and $\beta$, which can be seen as evidence of criticality. We numerically determine the various exponents which characterize this criticality.\end{abstract}

PACS numbers: 05.10.-a  71.45.Gm  89.65.Gh  05.45.Tp\\


The analysis of financial data by methods developed for physical systems, has extensively attracted the interest of physicists \cite{Mantegna2000l,Takayasu2002,Bunde2002,Bachelier1900,Mandelbrot1963,Pagan1996,Gabaix2003,Gopikrishnan98,Gopikrishnan99,Mantegna2000,Oswiecimka2005,Kwapien2005}.
In fact, financial markets are strongly fluctuating complex systems whose dynamics are difficult to understand because of the complexity of their internal elements and correlations, and also because of the many intractable external factors acting on them.
However, remarkably enough, the interactions between these various ingredients generate many observables whose statistical properties  appear to be similar for quite different markets.  Consequently, we are allowed to refer to some ``universal'' trends, on which we focus herein.

As a matter of fact, it has been observed,  in financial data, that rare events give raise to pronounced tails in the appropriate probability distributions --- tails that are in fact frequently found in complex systems. Such is the case of the return distributions associated with time series \cite{Mandelbrot1963,Mandelbrot1997} on varying time scales, $\Delta t$.
These fat tails  reveal long-range correlations that frequently cause standard statistical mechanics to be inadequate for describing them.
This kind of scenario also emerges in the systems that  permanently reside in the neighborhood of their critical point, where physical quantities present power-law dependences of the type $f(x)\sim x^{-\tau}$, characterized by a critical exponent $\tau$.

   Nonextensive statistical mechanics \cite{Tsallis1988,Tsallis2009},
 a current generalization of the Boltzmann-Gibbs (BG) statistical mechanics when  its associated entropy $S_{BG}$ does not obey the standard asymptotic behavior $S_{BG}(N)\sim N$ for $N\to \infty$ ($N$ being the number of elements),
 occurs useful in the description of such complex finantial systems \cite{Tsallis2003,Fredic2003,Rak2007,Drozdz2007,Ludescher2011,Tsallis16}.

 This theory has been developed around the concept of nonadditive entropy, which is maximized, with the appropriate constraints \cite{TsalLevy95}, by the family of the $q$-Gaussian distributions
 \begin{equation}
 \label{eq.1}
  G_q(x)= A(q,\beta_q) \exp_q[-\beta_q(x-\mu_q)^2],
 \end{equation}
 where $q$ is a characteristic index, $\beta_q$ is a sort of inverse  temperature \cite{Tsallis1998}, $\mu_q$ is the Escort averaged fist moment \cite{Estrada2009}, $A(q,\beta)$ is a normalization factor,   and the function
 \begin{equation}
 \label{eq.2}
 \exp_q(x)\equiv[1+(1-q)x]_+^{1/(1-q)}
 \end{equation}
 with $[x]_+ = x$ if $x > 0$ and $[x]_+ = 0$ otherwise,
  is a generalization of the exponential function --- the $q\to 1$ limit makes  $\exp_{q}(x) \to \exp(x)$ ---. The normalization factor in eq.~(\ref{eq.1})        reads  \cite{Prato1999},  for the values of $q$ we are now involved ($1<q<3$):
 \begin{equation}
 \label{eq.3}
  A(q,\beta_q)=
\sqrt{\frac{q-1}{\pi}\beta_q}\frac{\Gamma\left [ {\frac{1}{q-1}} \right ]}{\Gamma\left [ {\frac{3-q}{2\left ( {q-1} \right )}} \right ]}. \end{equation}

The $q$-Gaussian distribution (\ref{eq.1})  generalizes the Gaussian distribution in a similar  way as  nonadditive entropy generalizes $S_{BG}$ \cite{Tsallis1988}. In fact,    the $q\to 1$ limit makes eqs.~(\ref{eq.1}-\ref{eq.3}) to recover   the Gaussian distribution, i.e., $G_1(x)=\frac{1}{\sigma_1\sqrt{2 \pi}} \, \exp[\frac{(x-\mu_1)^2}{2\sigma_1^2}]$.
A generalization of the Central Limit Theorem, where the $q$-Gaussian distributions themselves become their attractors, has already  been formulated \cite{Moyano06,Umarov08,Umarov16}.

In the present work, we follow along the lines of \cite{Kwapien2012}. Namely, we apply a nonextensive statistical  analysis to empirical distribution data of normalized returns in financial market.
Our objective is to uncover some  empirical laws that seem to govern such a financial markets.

The  absolute normalized returns are conventionally  defined in the following manner. For the time series $W(t)$ that represent the prizes --- or the  market index value --- at time $t$,
the {\it returns} over a sample interval $\Delta t$, $R_{\Delta t}(t)$, are defined as
\begin{equation}
\label{eq.4}
R_{\Delta t}(t)\equiv
\ln W(t+\Delta t)- \ln W(t)
\simeq\frac{W(t+\Delta t)-W(t)}{W(t)},
\end{equation}
where the approximation holds for small variations of $W(t)$. By centering and normalizing $R_{\Delta t}(t)$, to have unit variance, the {\it normalized} returns are  obtained:
\begin{equation}
\label{eq.5}
r_{\Delta t}(t)\equiv
[R_{\Delta t}(t)-\langle R_{\Delta t}(t) \rangle_T]/v_{\Delta t}
\end{equation}
where $\langle\dots \rangle_T$ denotes a time average, and volatility $v_{\Delta t}$ is the standard deviation of the returns over the period $T$.

In the spirit of fig.~61 in \cite{Kwapien2012}, we are interested in studding the  cumulative distribution functions (CDF) of the {\it absolute} normalized returns, for   different  time scales $\Delta t$, of the $100$ American companies with the highest market capitalization. In other words, we analyze the  probability for an absolute return to be larger than a threshold $x$, i.e.,  $CDF(x)\equiv P(|r_{\Delta t}|>x)$. The negative and positive wings of empirical distributions are supposed to present negligible quantitative discrepancies and, consequently, we focus on the analysis of absolute returns.
\begin{figure}
\begin{center}
\includegraphics[width=1.\textwidth,angle=0]{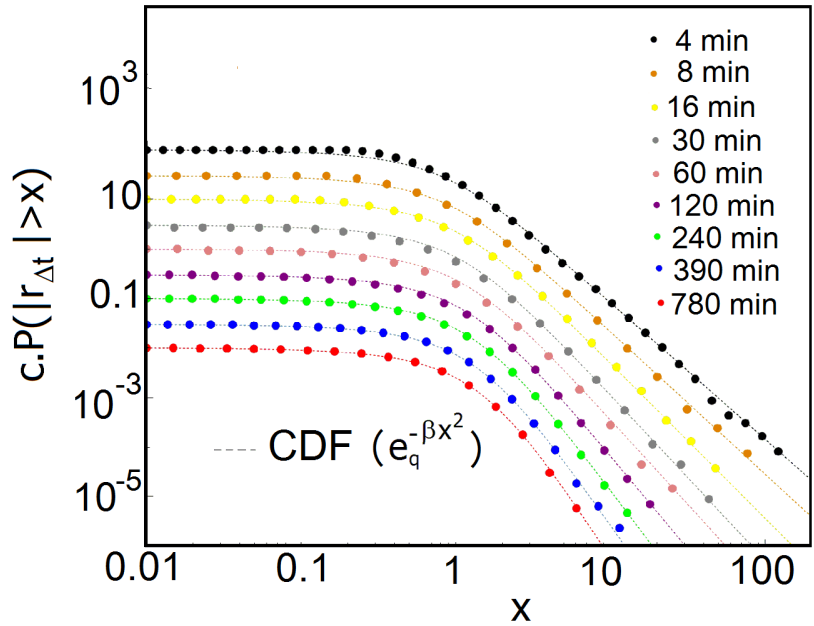}
\end{center}
\caption{(Color online) Cumulative distributions of absolute normalized returns that correspond to  different time scales $\Delta t$ for the 100 American companies with the highest market capitalization (points), and the fitted cumulative $q$-Gaussian distributions (lines). In order to better visualize the results, each $q$-Gaussian CDF and the respective experimental data have been multiplied by a positive factor, $c\ne 1$.}
\label{fig.1}
\end{figure}

 The asymptotic behavior of such normalized returns has been observed to follow an asymptotical power-law-like dependence of the type  $CDF(x) \sim 1/x^{\alpha}$ ($\alpha>0$). This is but one  of the arguments that make $q$-Gaussian distributions attractive to describe them; indeed    $q$-Gaussian asymptotically ($x\gg 1$) develop a power-law  form $G_q(x)\sim x^{2/1-q}$.

  First, we analytically obtain the CDF of a  $q$-Gaussian probability distribution function (pdf), with $\mu=0$ and  re-normalized   temperature $\beta^{-1}$,  as
\begin{equation}
\label{eq.6}
 P(\left| r_{\Delta t}\right| > x) =1-2A(q,\beta) \, \, x  \, \, _{2}F_{1}\left(\alpha, \delta;\gamma, \tau\right),
\end{equation}
where $\alpha= 1/2 $, $\delta=1/(q-1)$, $\gamma=3/2$,  $\tau=\beta\left ( 1-q \right )x^{2}$ and where $_{2}F_1\left(\alpha, \delta;\gamma, \tau\right)$ is the hypergeometric function. Eq.~(\ref{eq.6}) provides  a $q$-dependent asymptotical ($x\gg 1$) behavior  of the type $\sim x^{(q-3)/(q-1)}$, that  fits  the $\alpha$-dependent asymptotical behavior of absolute normalized returns, and provides  the $q$-Gaussian index through the relation:
\begin{equation}
\label{eq.7}
q=\frac{3+\alpha}{1+\alpha}.
\end{equation}

Even in the case that  empirical data of a particular time scale  did not attain the asymptotical behavior yet, we observe that cumulative distributions are also properly fitted  by the CDF of a $q$-Gaussian pdf (\ref{eq.6}). We obtain the value of $\beta$ associated  to the  index $q$ that corresponds to each time scale $\Delta t$, by a least squares fitting technique. Our $q$ versus $\Delta t$ results (see table~\ref{tab.1}), are in a quite satisfactory  agreement  with \cite{Kwapien2012}.
\begin{table}[h]
\caption{Time scales of absolute normalized returns, and the  $(q,\beta)$ values of their respective estimated CDFs.}
\label{tab.1}
\begin{center}
 \begin{tabular}{lcr}
$\Delta t$ &  $q$ & $\beta$  \\
\noalign{\smallskip}\hline\noalign{\smallskip}
4 & 1.53 &  1.78\\
8 & 1.52 &  1.67\\
 16& 1.48 &  1.52\\
 30 & 1.46 & 1.42 \\
 60 &1.45 &  1.33\\
 120 & 1.42 & 1.25 \\
 240 & 1.39 & 1.14 \\
 390 & 1.37 & 1.10 \\
 780 & 1.35 & 1.03
\end{tabular}
\end{center}
\end{table}

\begin{figure}
\begin{center}
\includegraphics[width=1.\textwidth,angle=0]{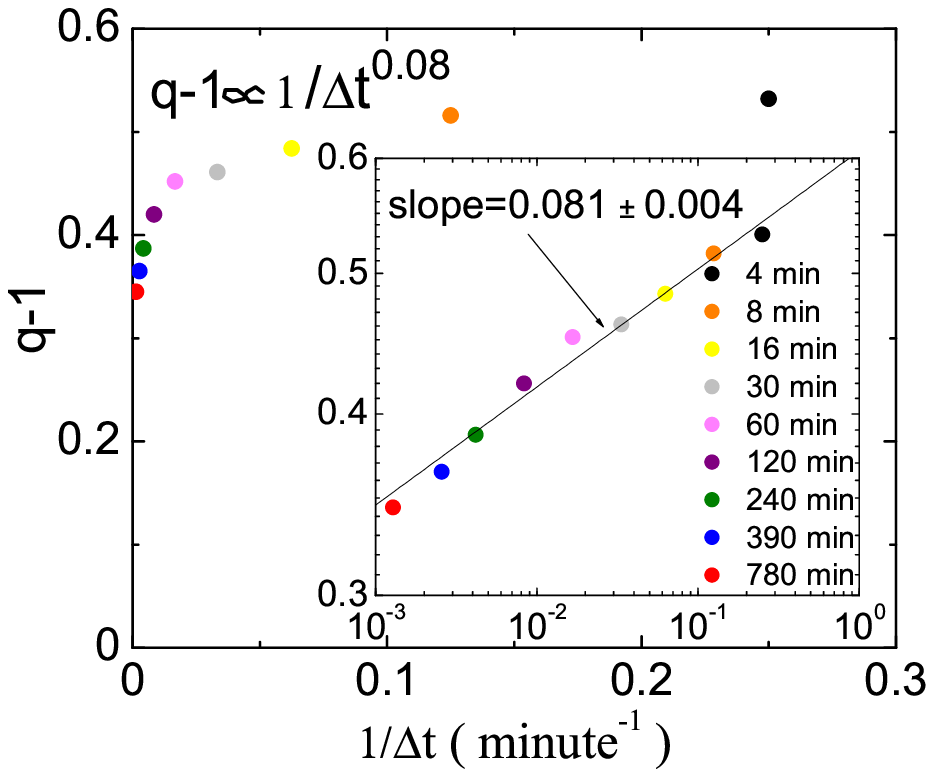}
\end{center}
\caption{(Color on line)  Dependence, versus temporal scale $\Delta t$,  of nonextensive index $q$, for the estimated $q$-Gaussian  pdfs of normalized absolute returns. Inset: Log-log representation shows a  power-law dependence of the type $q-1\propto \Delta t^{-\tau}$, with   $\tau= 0.081\pm 0.004$.  }
\label{fig.2}
\end{figure}

The fitted CDF  are represented, for all time scales, in fig.~\ref{fig.1}, together with the experimental data  provided in  \cite{Kwapien2012}.
The convergence of the CDF exhibits that, as $\Delta t$ increases, the value of $q$ decreases. Hypothetical final convergence to a Gaussian ($q\to 1$) appears to be abrupt (see fig.~\ref{fig.2}).

\begin{figure}
\begin{center}
\includegraphics[width=1.\textwidth,angle=0]{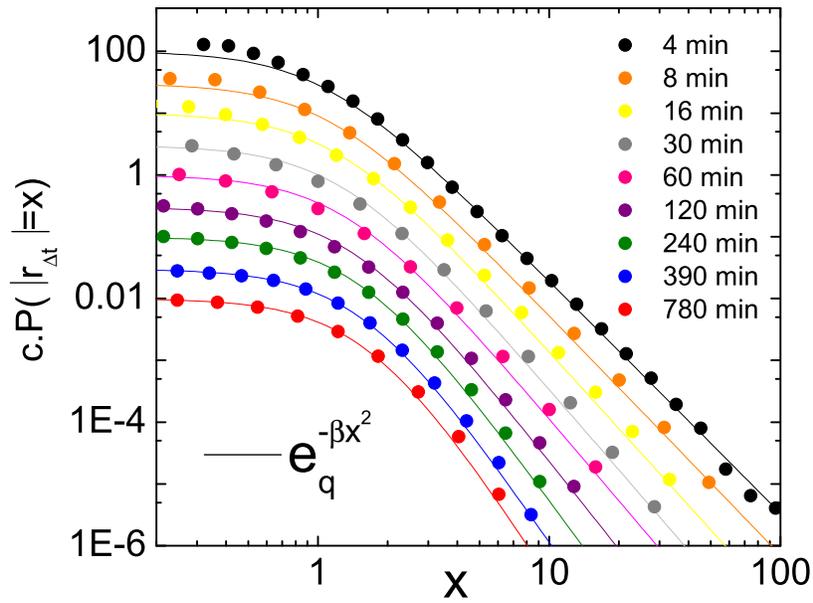}
\end{center}
\caption{(Color on line)  Log-log representation of the  probability density distributions of absolute normalized returns for different time scales $\Delta t$. Points have been obtained  by numerical derivation of  cumulative  values. Lines represent  the $q$-Gaussian pdfs that lead to the fitted CDFs  in fig.\ref{fig.1}. In order to better visualize the results, each $q$-Gaussian and the respective numerically estimated values,  are multiplied by a positive factor $c\ne 1$. }
\label{fig.3}
\end{figure}

\begin{figure}
\begin{center}
\includegraphics[width=1.\textwidth,angle=0]{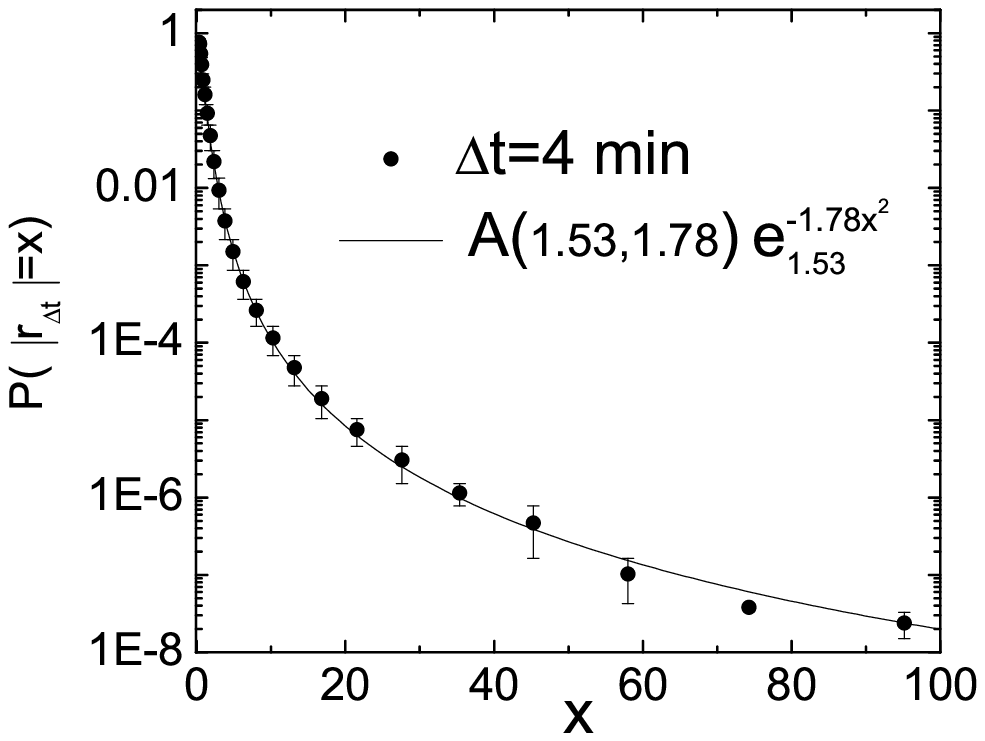}
\end{center}
\caption{Log-linear representation of the  probability density distribution of absolute normalized returns, for  time scale $\Delta t=4$. Points have been obtained  by numerical derivation of  cumulative  values.  Line represent  the $q$-Gaussian pdf that leads to the corresponding CDF  in fig.\ref{fig.1}, i.e.,  $q=1.53$ and $\beta=1.78$.  }
\label{fig.4}
\end{figure}
Fig.~\ref{fig.3} and fig.~\ref{fig.4} exhibit the good agreement with the $q$-Gaussian pdfs that lead to eq.~(\ref{eq.6}), with respect to  the derivatives of the experimental cumulative distributions.

\begin{figure}
\begin{center}
\includegraphics[width=1.\textwidth,angle=0]{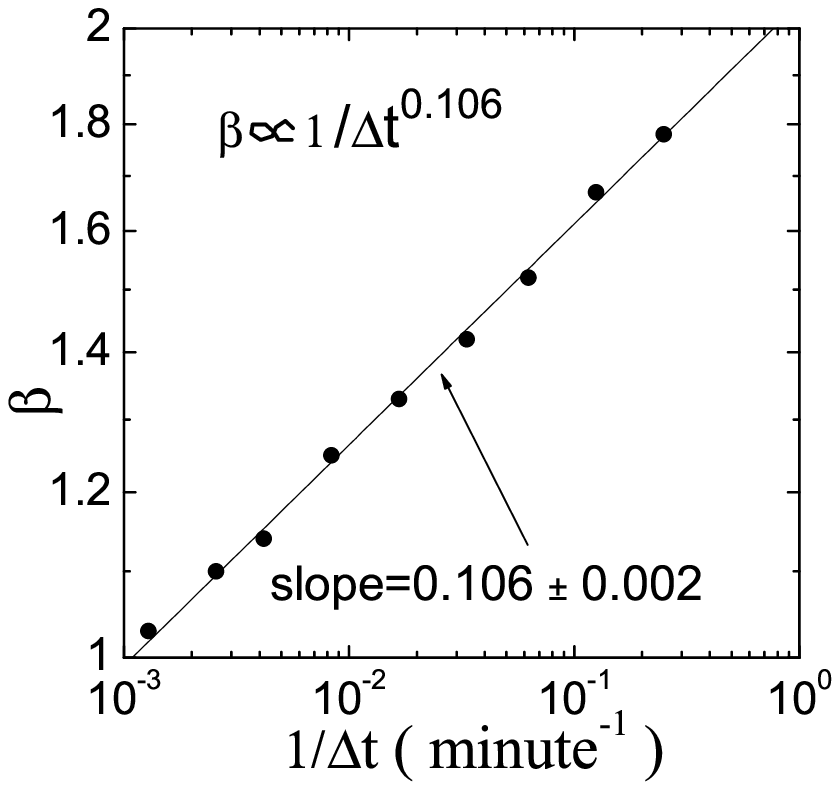}
\end{center}
\caption{Log-log representation of the   re-normalized inverse temperature $\beta$ versus temporal scale $\Delta t$,  for the estimated $q$-Gaussian  pdfs of normalized absolute returns. A  power-law dependence of the type $\beta^{-1}\propto \Delta t^{-\gamma}$ is observed, with   $\gamma= 0.106\pm 0.002$. }
\label{fig.5}
\end{figure}
\begin{figure}
\begin{center}
\includegraphics[width=1.\textwidth,angle=0]{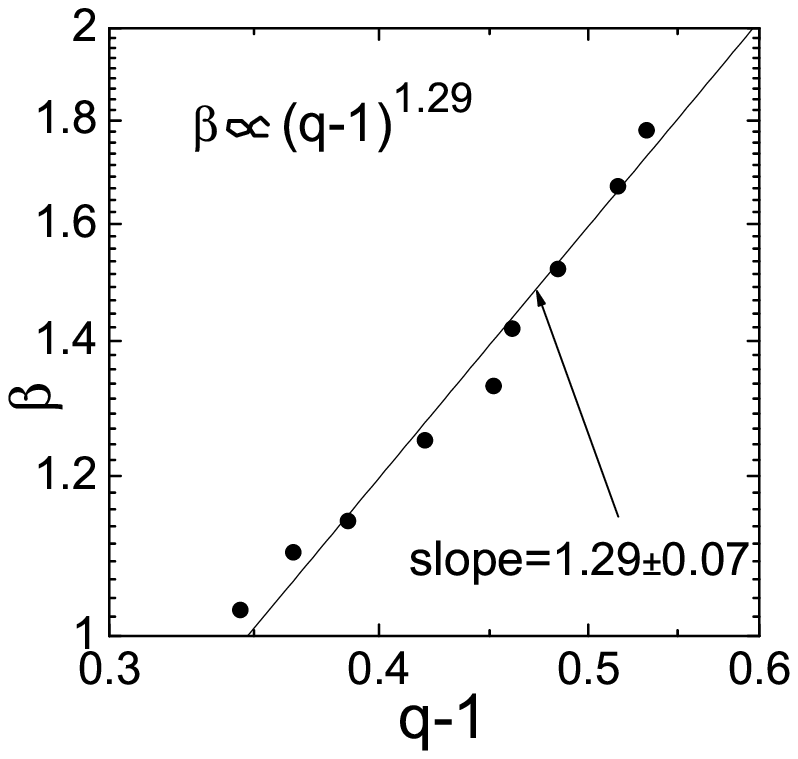}
\end{center}
\caption{Log-log representation of the re-normalized inverse temperature $\beta$ versus $q-1$, for the estimated $q$-Gaussian  pdfs of normalized absolute returns. A  power-law dependence of the type $\beta^{-1}\propto (q-1)^{-\delta}$ is observed, with   $\delta= 1.29\pm 0.07$. }
\label{fig.6}
\end{figure}

We have also observed that simple relations exist between the quantities $(q,\beta)$ involved for each  time scale $\Delta t$.
  A  power-law dependence  is observed for both $q$ and $\beta$  as a function of time scale $\Delta t$, the exponents being  $\tau=0.081\pm0.004$  (fig.~\ref{fig.2}) and $\gamma=0.106\pm 0.002 $  (fig.~\ref{fig.5}).
 Fig.~\ref{fig.6} shows in fact that the re-normalized  temperature $\beta^{-1}$  is not a free value, but it also exhibits a power-law dependence versus $q$, mainly   $\beta^{-1}=(q-1)^{-\delta}$, with   $\delta= 1.29\pm 0.07$.

Summarizing, our results show that $q$-statistics describes  complex systems that emerge in the analysis of the present  particular financial data. Similar results have been previously obtained  \cite{Tsallis2003,Rak2007,Kwapien2012}. But, undoubtedly, the  novelty of the present results is that we have also exhibited that both parameters $(q,\beta)$ of the nonextensive scenario  are specific values that are fixed by $\Delta t$. Such a behavior is analogous to the behavior of a variety of other  systems that are properly described by $q$-statistics, for example  scale-free $d$-dimensional geographically-located networks \cite{Brito2016}, quark-gluon soup in high-energy particle collisions \cite{Walton2000}, LHC/CERN and RHIC/Brookhaven experiments \cite{Khachatryan2010} and anomalous diffusion in confined granular media \cite{Granular2016}. Another simple and paradigmatic example is the logistic map where, as a reminiscence of this type of behavior, the  $q$-generalized Lyapunov exponent depends on the value of $q$ that characterizes the sensitivity to initial conditions at the edge of chaos  \cite{Lyra1998,Baldovin2004}. This frequent feature comes from the fact that $q$-statistics typically emerges at critical-like regimes and appears to be deeply related to an hierarchical occupation of phase space.

%

We are  grateful to professors J. Kwapie\'{n} and S. Dro\.{z}d\.{z} for sharing with us their empirical cumulative distribution data. One of us (G. R.) is grateful to professor C. Tsallis for is fruitful suggestions. One of us (G. R.) also acknowledges the warm hospitality at the CBPF (Brazil) and the partial financial support by the John Templeton Foundation (USA).


\begin{thebibliography}{99}

\bibitem{Mantegna2000l}
R. N. Mantegna and  H. E. Stanley,
{\it An introduction to Econophysics: Correlations and Complexity in Finance}
(Cambridge University Press, Cambridge, 2000).

\bibitem{Takayasu2002}
H. Takayasu H. (ed.)
{\it Empirical Science of Finantial Fluctuations: The Advent of Econophysics}
(Springer, Berlin, 2002).

\bibitem{Bunde2002}
A. Bunde,  H. J. Schellnhuber and J. Kropp,
{\it The Science of Disasters: Climate Disruptions, Hearth Attacks and Market Crashes }
(Springer, Berlin, 2002).


\bibitem{Bachelier1900}
L. Bachelier, Ann. Sci. \'{E}cole Norm. Suppl. {\bf 3}, 21 (1900).


\bibitem{Mandelbrot1963}
B. B. Mandelbrot, J. Business {\bf 36}, 294 (1962).

\bibitem{Pagan1996}
A. Pagan,
J. Empirical Finance {\bf 3}, 15 (1996).

\bibitem{Gabaix2003}
X. Gabaix, P. Gopikrishnan, V. Plerou and   H. E. Stanley,
Nature {\bf 423},  267 (2003).


\bibitem{Gopikrishnan98}
P. Gopikrishnan, M. Meyer, L. A. N.  Amaral, and H. E. Stanley, Eur. Phys. J. B {\bf 3}, 138 (1998).


\bibitem{Gopikrishnan99}
P. Gopikrishnan, V. Plerou, L. A. N. Amaral, M. Meyer and  H. E. Stanley,
 Phys. Rev. E {\bf 60},  5305 (1999).


\bibitem{Mantegna2000}
M. Denys, T. Gubiec, M. Jagielski, R. Kutner, and   H. E. Stanley,
Phys. Rev. E {\bf 94},  042305 (2016).


\bibitem{Oswiecimka2005}
P. O\'{s}wiecimka, J. Kwapie\'{n}  and S. Dro\.{z}o\.{z},
Physica A {\bf 347},  626 (2005).

\bibitem{Kwapien2005}
J. Kwapie\'{n}, P. O\'{s}wiecimka and S. Dro\.{z}o\.{z},
Physica A {\bf 350},  466 (2005).


\bibitem{Mandelbrot1997}
B. B. Mandelbrot,
{\it Fractals and Scaling in Finance} (Springer, Berlin, 1997).

\bibitem{Tsallis1988}
C. Tsallis,
J. Stat. Phys. {\bf 52},  479 (1988).

\bibitem{Tsallis2009}
C. Tsallis,
{\it Introduction to Nonextensive Statistical Mechanics -- Approaching a Complex World} (Springer, New York, 2009).

\bibitem{Tsallis2003}
C. Tsallis, C. Anteneodo,  L. Borland and R. Osorio,
Physica A {\bf 324},  89 (2003).


\bibitem{Fredic2003}
F. Michael and   M. D. Johnson,
Physica A {\bf 320}, 525 (2003).


\bibitem{Rak2007}
R. Rak, S. Dro\.{z}d\.{z} and J. Kwapie\'{n},
Physica A {\bf 374},  315 (2007).

\bibitem{Drozdz2007}
 S. Dro\.{z}d\.{z},   M. Forczek,  J. Kwapie\'{n}, P.  O\'{s}wiecimka  and   R. Rak,
Physica A {\bf 383},  59 (2007).

\bibitem{Ludescher2011}
J. Ludescher, C. Tsallis and A. Bunde,
Eur. Phys. Lett. {\bf 95},  68002 (2011).

\bibitem{Tsallis16}
 C. Tsallis,
Chaos, Solitons and Fractals  {\bf 88},  254 (2016).

\bibitem{TsalLevy95}
C. Tsallis, S. V. F. Levy, A. M. C. Souza and R. Maynard,
Phys. Rev. Lett. {\bf 75},  3589 (1995).



\bibitem{Tsallis1998}
C. Tsallis, R.S. Mendes  and  A. R. Plastino,
Physica A {\bf 261},  534 (1998).

\bibitem{Estrada2009}
C. Tsallis, A. R. Plastino and  R. F. Alvarez-Estrada,
 J. Math. Phys. {\bf 50}, 043303  (2009).

\bibitem{Prato1999}
D. Prato  and C. Tsallis,
Phys. Rev. E {\bf 60},  2398 (1999).

\bibitem{Moyano06}
L.G., Moyano, C. Tsallis and  M. Gell-Mann,
Europhys Lett {\bf 73},  813 (2006).

\bibitem{Umarov08}
S. Umarov, C. Tsallis and S. Steinberg,
Milan J. Math.  {\bf 76}, 307 (2008).

\bibitem{Umarov16}
S. Umarov and C. Tsallis,
J. Phys. A: Math. Theor.  {\bf 49}, 415204(2016).


\bibitem{Kwapien2012}
 J. Kwapie\'{n} and S. Dro\.{z}d\.{z},
Physics Report {\bf 515},  115 (2012).


\bibitem{Brito2016}
S. G. A. Brito, L. R. da Silva and C. Tsallis,
Nature - Scientific  Reports {\bf 6},  27992 (2016).


\bibitem{Walton2000}
D. B.  Walton and Rafelshi J.,
Phys  Rev. Lett. {\bf 84},  31 (2000).


\bibitem{Khachatryan2010}
V. Khachatryan et al.,
Phys. Rev. Lett. {\bf 105},  022002 (2010).

\bibitem{Granular2016}
G. Combe, V. Richefeu, M. Stasiak and  A. P. F. Atman,
Phys  Rev. Lett. {\bf 115}, 238301 (2016).

\bibitem{Lyra1998}
M. L. Lyra and C. Tsallis,
Phys. Rev. Lett {\bf 80},  53 (1998).

\bibitem{Baldovin2004}
F. Baldovin and A. Robledo,
Phys. Rev. E {\bf 69},  045202(R) (2004).

\end{thebibliography}
\end{document}